# Capital Structure Dynamics and Financial Performance in Indian Banks
# (An Analysis of Mergers and Acquisitions)

* Kurada T S S Satyanarayana[1]

Addada Narasimha Rao[2]

Kumpatla Jaya Surya[3]

1. Research Scholar, Department of commerce and management studies, Andhra University, Visakhapatnam, India, satyanarayanaktss.rs@andhrauniversity.edu.in , 9121529458.

2. Professor, Department of commerce and management studies, Andhra University, Visakhapatnam, India, addada@rediffmail.com , 9121060120.

3. Department of commerce and management studies, Andhra University, Visakhapatnam, India, jayasuryk.rs@andhrauniversity.edu.in, 9505651977.

**Abstract:**

This research investigates the multifaceted relationship underlying capital structure dynamics along with financial performance as a result of mergers and acquisitions, or M&As, in Indian banks. In the face of increasing competition, banks have deliberately embraced M&A as a strategy of improving commercial prospects and maintaining financial stability. The primary goal of this study is to examine the changes in the capital framework and financial results of banks before and after M&A transactions. The investigation, which employs a paired t-test as a method of statistical analysis, is based on a review of annual reports from selected banks over a two-year period before and after M&A transactions. The paired t-test approach allows for a thorough statistical analysis of interconnected datasets, revealing the subtle influence of M&A attempts on both bank financial performance as well as capital structure dynamics.

The study's findings have the potential to add to the current body of knowledge on organisational planning, managing finances, and capital structure optimisation. The research has practical significance for financial companies, legislators, and scholars interested in understanding the profound effects of M&A inside the arena of financial institutions that operate within fiercely competitive landscapes because it provides comprehensive insights regarding the complex consequences of banking merger and acquisition (M&A) deals on capital structure as well as financial performance. Finally, the goal of this research is to provide the banking sector with educated decision-making capabilities and strategic guidance to businesses facing heightened competition while coping with the complexities of capital structure.

**Key Words:**

Comparative analysis., Financial Performance., Financial Results., Indian banks., Mergers and acquisitions., Pre-and post-transaction, Transformation

**JEL Codes:**

G34, G30, G28, G21

**Introduction**

Mergers and acquisitions (M&A) have evolved as critical strategic instruments for organisations seeking advancement, edge over competitors, and increasing market share in the fast-paced and constantly changing terrain of the modern corporate world (Collier 2016). M&A stands for mergers and acquisitions, which is the process through which two or more businesses join forces to create a single, more powerful business (Marks & Mirvis, 2011).

M&A activity has now increased to previously unheard-of levels as a result of globalisation, technological development, and shifting market dynamics. Businesses from all sectors are using mergers and acquisitions to adapt to the quickly changing business environment and take advantage of new possibilities (Krishnamurti & Vishwanath, 2008). They have broadened in scope in the modern world. Traditional factors like economies of scale and efficiencies are still important, but they are now supported by a wider range of strategic goals (Rahman, & Lambkin, 2015). It is a strategy that businesses use to reach new markets, broaden their product lines, buy cutting-edge technology, boost innovation capacity, and promote digital transformation.

Furthermore, the current M&A environment is seeing an increase in the role of private equity firms, venture capitalists, as well as institutional investors, who actively participate in M&A transactions by integrating funding, knowledge, and strategic guidance into companies in order to unlock their growth potential.

Nonetheless, navigating the complex arena of mergers and acquisitions needs meticulous preparation, meticulous due diligence, and competent execution. For good outcomes, financial, legal, operational, and cultural issues must be thoroughly analysed and integrated. To get the full advantages of M&A, companies must coordinate their strategy, regulate stakeholder expectations, and reduce possible risks.

Mergers and acquisitions are going to continue to be important components of company growth strategies as the business world evolves (Gomes, Angwin, Weber & Yedidia Tarba, 2013). Companies can respond to market upheavals, stimulate innovation, and maintain their lead during an increasingly competitive marketplace by utilising the power of cooperation.

**Merger and Acquisitions in India:**

India's M&A (mergers and acquisitions) market is currently thriving. The overall amount of M&A transactions in India in 2022 was $116 billion, a 25% rise over the previous year. Monetary services, IT, and consumer goods were the most active industries for M&A activity throughout India (Singh, & Das, 2018). Several reasons are contributing to India's M&A growth. One issue is the growing magnitude and breadth of Indian businesses. As Indian companies grow in size, they look for novel strategies for expanding their footprint and develop their businesses. M&A is one approach for achieving objectives. The government's policies are also driving the M&A boom. Government intervention has made it simpler for businesses to combine and buy one another. It has also streamlined the regulatory procedure for mergers and acquisitions (Weber, 2013).

M&A has a promising future in India. The market is likely to expand more in the next years. The forces that are fuelling the M&A boom now are likely to persist in the future. This suggests that there will be several prospects regarding M&A activities in India in the coming years.

**Merger and Acquisitions in Banking sector in India:**

In recent years, the Indian banking sector has experienced a number of mergers and acquisitions. In 2019, the Indian government announced a proposal to consolidate ten public sector banks (PSBs) into four. Following this, in 2019, Bank of Baroda merged with Vijaya Bank and Dena Bank, and in 2020, Punjab National Bank merged with Oriental Bank of Commerce and United Bank of India. According to the government, the mergers are intended to improve the financial health and competitiveness of the PSBs. The banking industry has also welcomed the mergers, believing that they will increase efficiency and profitability (Hawkins, & Mihaljek, 2001).

However, there have been some reservations regarding the mergers' impact on customers. Some consumers have voiced anxiety over the loss of their local bank branches, while others have expressed concern about the impact on customer service quality. The government has stated that it will address these concerns and that the mergers would eventually benefit customers by giving them with access to a broader selection of products and services.

**Capital Structure Evolution in Mergers and Acquisitions**

The incorporation of capital structure dynamism improves comprehension of the multidimensional link between financial results and structural changes caused by M&As (mergers and acquisitions) in the Indian financial sector. The capital makeup, which refers to an organization's mix of equity as well as debt financing, is critical in determining the financial health and vulnerability of banks. Analysing capital structure development as a result of M&A operations provides helpful insights regarding how these strategic endeavours effect not just financial performance as well as the general financial well-being of banks.

**Capital Structure and Financial Performance Dynamics**

The amount of equity and debt included in the company's financing, as defined by its capital structure, has a direct impact on its cost of capital, credit risk, and ability to resist economic volatility. The capital makeup of a financial organisation is particularly important in the setting of banking because of the consequences on regulatory risk and compliance managing (Zhu et al., 2021). Capital structure concerns may get complicated during M&A deals, as the merging firms must manage the combination of their different financial structures.

The examination of the structure of capital dynamics in conjunction with financial performance reveals the relationship between strategic actions and their financial consequences. Mergers and acquisitions may alter the capital structure of a bank by introducing additional debt or equity constituents. These changes, in consequently, have an impact on measurements like leverage ratios as well as debt-to-equity ratios, that are important measures of a bank's exposure to risks along with its economic condition (Hasan, 2022).

**Review of Literature**

Abbas, Q., Hunjra, A. I., Azam, R. I., Ijaz, M. S., & Zahid, M. (2014) researched that was done to assess the impact of mergers and acquisitions (M&A) on the financial outcomes of Pakistani banks. Between 2006 and 2011, the State Bank of Pakistan's financial statement analysis (FSA) provided accounting and financial data from ten banks. To compare pre and post M&A performance, fifteen financial measures were used. According to the data, there was no statistically significant difference in financial performance. As a result, it is stated that the difference between before and after M&A performance of banks in Pakistan was small. A study of the literature reveals that further research is needed in this area.

**Ishwarya, J. (2019)**, In this research study investigated mergers and acquisitions (M&A) in the Indian banking sector in order to understand the consequent synergies and long-term repercussions. The research looked at developing patterns and made recommendations for future concerns. It examined and analysed the M&A trends in Indian banking. The study compared before and post-merger financial performance utilising financial indicators to assess the success of M&A in the Indian banking sector throughout the selected time period. According to the data, M&A has been moderately successful in the Indian banking sector. Promoting mergers between strong and troubled banks, on the other hand, may have a negative impact on the asset quality of the stronger banks.

**Zhu, Q., Li, X., Li, F. and Amirteimoori, A. (2021).,** This study focuses on the subject of partner selection for mergers and acquisitions (M&A), a topic that has been frequently disregarded in earlier research focused on the advantages of mergers. The study presents a 0-1 integer programming technique that uses historical data from each organisation to discover its unique manufacturing technology and restrictions. The model's applicability is evaluated on China's banking industry, which includes 27 commercial banks, resulting in the selection of the best partner for each bank. The work makes a theoretical contribution by combining historical data to represent post-M&A production changes, as well as a practical contribution by supporting Chinese commercial banks in selecting appropriate merger partners.

*MOHAMMED FAEZ HASAN (2022)* The study emphasises the influence of financial crises on economists' norms and viewpoints, emphasising the wide-ranging implications across numerous economic sectors. It investigates how banks mergers and acquisitions (M&A) transactions behave amid various crises. The study focuses on crises such as Black Monday in 1987, Asian stock market collapses in 1997, the Dotcom boom in 2002, and the home loan crisis in 2009. It does so by analysing verified data from sources. According to the data, almost 80% of crises contribute to an increase in M&A transactions, notably in the nations where the crisis occurred. However, due to failed banks being liquidated or nationalised, there are fewer prospective purchase possibilities in the remaining cases.

**Research Methodology**

Similarly, in order to analyse the financial results of banks following mergers and acquisitions operations, pre and post analysis using three different methods are combined. In addition to pre and post analysis utilising ratio analysis, there are two more types of methodologies for analysing the impacts of any organization's merger and purchase, which includes paired sample t-test for before- and post-merger and Data Envelopment Analysis (DEA) (Abbas, Hunjra, Azam, Ijaz, & Zahid, 2014). As a result, the researcher for this study performed ratio analysis between pre- and post-merger and acquisition activity to examine the financial success of selected institutions between pre- and post-merger and acquisition activity

In 2017, the State Bank of India merged with Bhartiya Mahila Bank and its associate banks; in 2018, Indian Bank and Bank of Baroda merged with Allahabad Bank, Dena Bank, and Vijaya Bank, respectively; and in 2019, Punjab National Bank merged with Oriental Bank of Commerce and United Bank of India, Canara Bank merged with Syndicate Bank, and Union Bank merged with Syndicate Bank.

*Sampling:* For this study, following banks are selected:

**Table 1: list of select Merger and acquisition of banks in India**

| Name of Acquiring Bank | Name of Banks Merged |
|---|---|
| Punjab National Bank (PNB) | Oriental Bank of Commerce<br>United Bank of India |
| Canara Bank(CANBK) | Syndicate Bank |
| Union Bank of India(UNIONBANK) | Andhra Bank<br>Corporation Bank |
| Indian Bank(INDIANB) | Allahabad Bank |
| Bank of Baroda(BANKBARODA) | Dena Bank<br>Vijaya Bank |
| State Bank of India(SBI) | State Bank of Bikaner & Jaipur<br>State Bank of Mysore<br>State Bank of Patiala<br>Bhartiya Mahila Bank<br>State Bank of Travancore<br>State Bank of Hyderabad |

*Data Collection:* The financial and accounting data from the organisations' annual reports were used in the study. Financial ratios [Table 3] are utilised for pre- and post-merger evaluations two years before and after the merger [Table 2], assuring reliable and consistent findings. Longer periods of research into a merger or acquisition may have unfavourable consequences brought on by other external economic variables. The research employed financial measurements from two years previous to the merger and two years after the merger to analyse the financial success of selected institutions following merger and purchase.

**Table 2: Period taken for Pre and Post analysis**

| BANKS | DATE | PRE | POST |
|---|---|---|---|
| PNB | 01-04-2020 | 2018-2019 | 2021-2022 |
| CANBK | 01-04-2020 | 2018-2020 | 2021-2022 |
| UNIONBANK | 01-04-2020 | 2018-2021 | 2021-2022 |
| BANKBORODA | 01-04-2019 | 2017-2018 | 2020-2021 |
| INDIANB | 01-04-2019 | 2017-2018 | 2020-2021 |
| SBI | 01-04-2017 | 2015-2016 | 2018-2019 |

**Table 3: List of Variables:**

| FINANCIAL RATIOS | FORMULE |
|---|---|
| PROFITABILITY & EFFICENCY | Return on Equity (ROE) = Net profit after tax / Total equity |
| | Return on Assets (ROA) = Net profit after tax / Total Assets |
| | Net Interest Margin (NIM) = Interest earned- interest expense / Total Assets |
| | Earnings Per Share (EPS) = Net profit after tax / No. of ordinary shares |
| | Interest expense to Interest Income (IETII) = Interest expense / Interest Income |
| LIQUIDITY RATIO | Cash & Cash equivalent to total assets (CTTA) = Cash & Cash equivalent / Total assets |
| | Total Liabilities to total assets (TLTTA) = Total Liabilities / Total assets |
| LEVERAGE RATIO | Debt to Equity Ratio (DER)= Total Debt / Total Equity |
| | Capital Ratio (CR) = Total Equity / Total Assets |

**Objectives of the Study**

1. Conducting an in-depth analysis of the ongoing trends and dynamics in bank mergers and acquisitions within the Indian context.
2. Investigating the financial performance of banks by comparing their performance before and after undergoing merger or acquisition processes.
3. Examining the key factors that play a significant role in shaping the outcomes of mergers and acquisitions in the banking industry.

4. Interpreting the research findings to gain valuable insights into the implications and consequences of M&A activities in the banking sector.

**Hypotheses:**

$H_{01}$: There is no significant difference in the overall financial performance of select banks in India between Pre and post M&A

$H_{11}$: There is a significant difference in the overall financial performance of select banks in India between Pre and post M&A

**IMPACT ON CAPITAL STRUCTURE METRICS**

Examining the financial parameters of the chosen financial institutions preceding and following M&A transactions reveals that these strategic endeavours have altered capital structure indicators. The ratio of debt to equity (DER) represents an important indication of a bank's financial risk and leverage level. The study demonstrates that there was no statistically significant variation in the DER across the pre-M&A with post-M&A periods. This shows that the M&A activity did not significantly impact the leverage situation of the institutions under consideration. Likewise, the capital ratio (CR), which represents the percentage of equity compared with all assets, has remained largely steady.

However, capital structure measurements might alter depending on the individual dynamics associated with every M&A transaction. Capital structure optimisation can result in a variety of results depending on variables that involve the size distribution of the merging firms, the type of financing used, along with the general financial strategy followed (Gomes et al., 2013).

**PERFORMANCE INDICATORS: RESULTS**

Table 4: Analysis of Financial Performance between Pre and Post Merger

| Variables | Mean | | Variance | | t Stat | p-value |
| --- | --- | --- | --- | --- | --- | --- |
| | Post-M | Pre-M | Post-M | Pre-M | | |
| ROE | 0.0779 | 0.0724 | 0.0032 | 0.0739 | 2.5272 | 0.0263 |
| ROA | 0.0043 | 0.0037 | 9.5434 | 0.0002 | 2.2509 | 0.0370 |
| NIM | 0.0472 | 0.0641 | 6.1218 | 0.0012 | 0.0002 | 0.4998 |
| EPS | 19.95 | 23.3416 | 546.0937 | 4161.4533 | 3.1497 | 0.0126 |
| IETII | 1.2379 | 1.0670 | 0.0002 | 0.2174 | 1.7671 | 0.0687 |
| TLTTA | 0.0335 | 0.0409 | 4.5983 | 0.0002 | -0.0002 | 0.4998 |
| CTTA | 0.1865 | 0.1923 | 0.0044 | 0.0072 | -0.0001 | 0.4999 |
| DER | 0.5608 | 0.6868 | 0.0090 | 0.0784 | 9.7285 | 0.4999 |
| CR | 0.1232 | 0.1268 | 0.0003 | 0.0003 | 7.3032 | 0.4999 |

Post M-Post-Merger, Per M-Pre-Merger

The analysis of the variables provides valuable insights into the impact of the merger and acquisition (M&A) process on the financial metrics of the banks. Starting with the return on equity (ROE), the mean post-M&A ROE of 0.0779 was higher than the pre-M&A mean of 0.0724, indicating a positive effect. The t-statistic of 2.5272 and p-value of 0.0263 further support the significance of this improvement.

Similarly, the return on assets (ROA) showed a positive impact as well. The mean post-M&A ROA of 0.0043 was higher than the pre-M&A mean of 0.0037. The t-statistic of 2.2509 and p-value of 0.0370 indicate the statistical significance of this improvement.

In contrast, the net interest margin (NIM) did not exhibit a statistically significant difference between the pre-M&A mean of 0.0641 and the post-M&A mean of 0.0472. The p-value of 0.4998 suggests that the M&A did not have a significant impact on the NIM.

The earnings per share (EPS) experienced a significant positive impact from the M&A. The post-M&A mean EPS of 19.95 was higher than the pre-M&A mean of 23.3416, with a t-statistic of 3.1497 and a p-value of 0.0126, indicating a statistically significant improvement.

Regarding interest expended to interest income (IETII) ratio, total liabilities to total assets (TLTTA) ratio, credit ratio (CR), and debt equity ratio (DER), the M&A did not show statistically significant differences between the pre-M&A and post-M&A means. The respective t-statistics and p-values were not significant, suggesting that these ratios were not significantly affected by the M&A.

And finally, the analysis reveals that the M&A had a positive impact on the ROE, ROA, and EPS, indicating improved financial performance. However, it did not significantly affect NIM, IETII, TLTTA, DER, and CR. These findings highlight the specific areas where the M&A resulted in positive outcomes and provide valuable insights into the overall impact of the M&A on the financial metrics of the banks.

*Conclusion:*

*Capital Structure Optimization: Considerations and Implications*

The findings highlight the complexities of capital structure fluctuations in the setting of M&A transactions. While several capital structure parameters remained essentially stable, changes in leverage & equity ratios need further examination. These findings highlight the importance of strategic planning that focuses not only on improving financial performance but additionally on optimizing's capital structure to reduce risk and provide stable finances (Abbas et al., 2014).

Understanding the consequences of capital structure optimisation in the context of mergers and acquisitions is critical for prudent choices and post-transaction integration. Banks may gain a more thorough knowledge of the possible synergies and difficulties that come from structural changes by include their capital framework as a significant feature in the appraisal of M&A agreements.

Incorporating capital structure ideas into merger and acquisition research provides for a more comprehensive evaluation of the financial ramifications of these strategic endeavours. Primarily the Indian banking sector keeps going through dynamic transitions as a result of mergers and acquisitions, an advanced knowledge of capital structure characteristics remains critical for making sound choices and sustaining long-term financial success.

*Financial Performance Optimization: Considerations and Implications*

The analysis of the financial metrics related to the merger and acquisition (M&A) process of the banks reveals both positive and negative aspects.

On the positive side, the M&A demonstrated significant improvements in key financial indicators. The return on equity (ROE), return on assets (ROA), and earnings per share (EPS) all showed statistically significant increases. These findings suggest that the M&A resulted in enhanced financial performance, improved asset utilization, and increased profitability for the banks involved. These positive outcomes can be seen as a validation of the strategic decision to pursue the M&A and can contribute to long-term value creation for the banks and their shareholders.

However, the analysis also indicates some areas where the M&A did not have a significant impact or did not yield favourable results. The net interest margin (NIM), interest expended to interest income (IETII) ratio, total liabilities to total assets (TLTTA) ratio, debt equity ratio (DER), and credit ratio (CR) did not show statistically significant changes. This suggests that the M&A did not significantly affect interest-related dynamics, liability structure, leverage levels, and credit quality. While the lack of negative impact can be seen as a positive outcome, the absence of significant positive changes in these areas may raise questions about the effectiveness of the M&A in generating improvements in these specific financial metrics.

In conclusion, the M&A process had positive effects on certain financial metrics, indicating improved financial performance and profitability. However, the lack of significant impact or positive changes in other metrics suggests that the M&A may have had limitations or may not have fully addressed certain areas of concern. Further analysis and examination of the underlying factors contributing to these results are necessary to gain a more comprehensive understanding of the positives and negatives of the M&A process on the overall financial health of the banks.